\documentclass[11pt,a4paper]{article}
\usepackage[hyperref]{acl2020}
\usepackage{preamble}

\usepackage{microtype}

\usepackage{amsmath}
\usepackage{cleveref}

\usepackage{amsfonts}
\usepackage{xcolor,colortbl}

\newcommand{\TODOh}[1]{}

\usepackage{listings}
\usepackage[utf8x]{inputenc}
\AtBeginDocument{\setlength\abovedisplayskip{5pt}}
\AtBeginDocument{\setlength\belowdisplayskip{5pt}}

\usepackage{tabu}

\aclfinalcopy 

\title{Align-Refine: Non-Autoregressive Speech Recognition \\ via Iterative Realignment}

\author{Ethan A. Chi \\
  Stanford University\thanks{ 
    \; Work done during an internship at Amazon AWS AI.
  } \\
  \texttt{ethanchi@cs.stanford.edu} \\\And
  Julian Salazar \\
  Amazon AWS AI \\
  \texttt{julsal@amazon.com} \\\And
  Katrin Kirchhoff \\
  Amazon AWS AI \\
  \texttt{katrinki@amazon.com} \\}

\date{}

\begin{document}
\maketitle
\begin{abstract}
Non-autoregressive models greatly improve decoding speed over typical sequence-to-sequence models,
but suffer from degraded performance.
Infilling and iterative refinement models make up some of this gap by editing the outputs of a non-autoregressive model,
but are constrained in the edits that they can make.
We propose \textit{iterative realignment}, where refinements occur over \textit{latent alignments} rather than output sequence space.
We demonstrate this in speech recognition with Align-Refine, an end-to-end Transformer-based model which refines connectionist temporal classification (CTC) alignments to allow length-changing insertions and deletions.
Align-Refine outperforms Imputer and Mask-CTC, matching an autoregressive baseline on WSJ at 1/14th the real-time factor and attaining a LibriSpeech test-other WER of 9.0\% without an LM. Our model is strong even in one iteration with a shallower decoder.
\end{abstract}

\section{Introduction}
Transformer encoder-decoder models \cite{vaswani2017attention} have achieved high performance in sequence-to-sequence tasks like neural machine translation (NMT; \citealp{edunov2018understanding}) and end-to-end automatic speech recognition (ASR; \citealp{karita2019comparative}), outperforming their RNN-based predecessors \cite{bahdanau2014neural, chan2016listen} while reducing training time via the non-recurrent self-attention mechanism.
However, these models remain \textit{autoregressive} during inference: tokens are generated one by one, with each token conditioned on all previous tokens.
Since this cannot be parallelized, decoding time is linear in the output sequence length, which is slow for long sequences and large decoders as in the Transformer.

By contrast, non-autoregressive models decode all output tokens independently and in parallel. When combined with self-attention, one gets fast, constant-time inference in NMT \cite{gu2017non} and end-to-end ASR \cite{salazar2019sanctc}.
However, these models underperform their autoregressive counterparts as the conditional independence between output tokens can give globally inconsistent outputs.\footnote{\citeauthor{gu2017non}\ call this \textit{the multimodality problem} as it is induced by the highly multimodal distribution of target translations (or in the case of ASR, frame-level alignments).}

\begin{figure}[t!]
    \centering
    \includegraphics[width=0.85\linewidth,trim={0 1.6cm 0 0},clip]{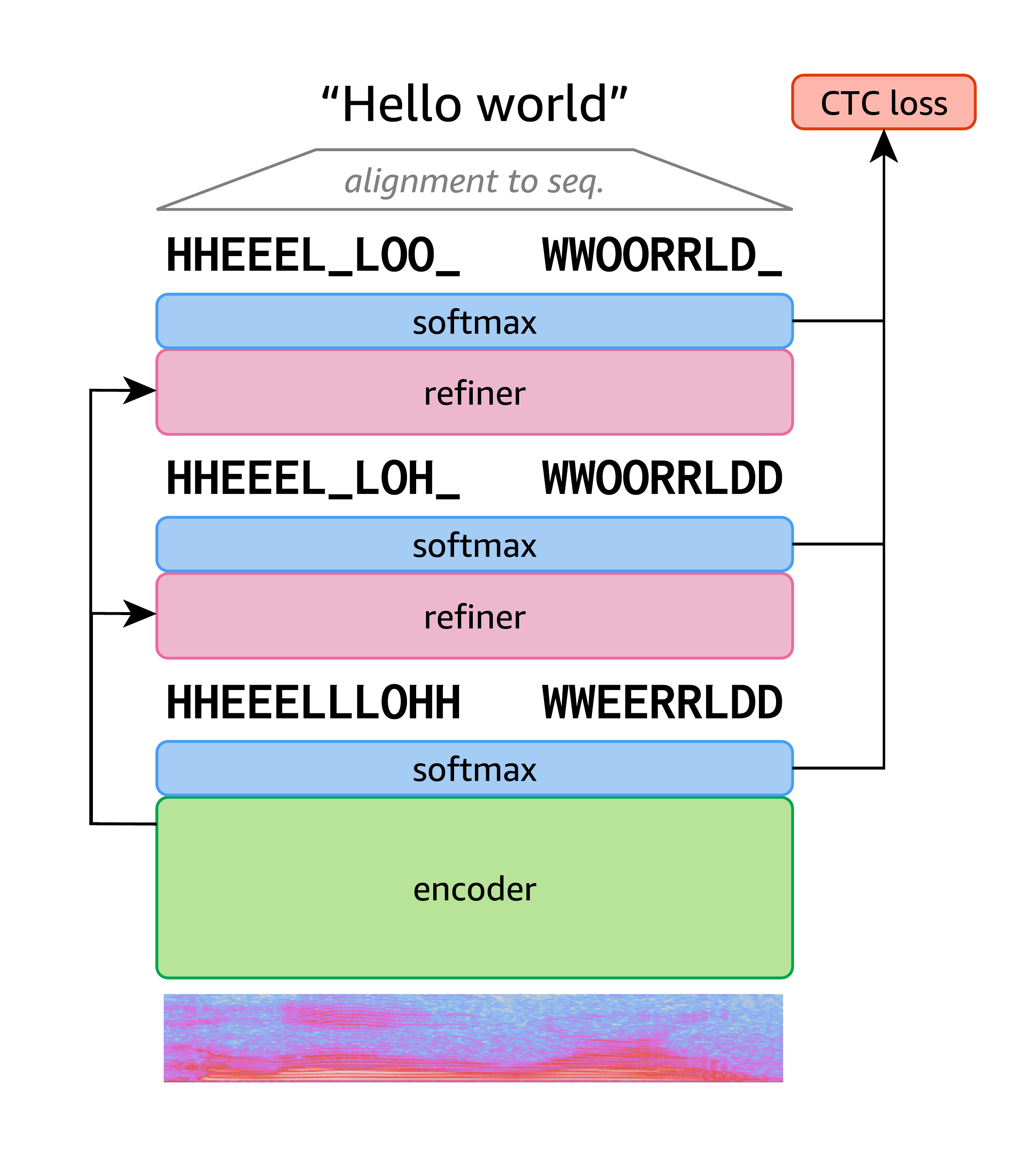}
    \caption{A Transformer encoder labels each input frame to give a latent alignment. The refiner, a non-causal Transformer decoder, conditions on this and the encoder to improve the alignment. After a bounded \# of iterations, the result is collapsed into the output.}
    \vspace{-15pt}
    \label{fig:align-refine}
\end{figure}

To mitigate these issues, Mask-Predict \citep{ghazvininejad2019mask} \textit{infills} output tokens in a fixed number of decoding passes, with a shrinking proportion of tokens being masked and predicted each time. Infilling is the prevailing non-autoregressive iterative decoding method in ASR, such as in A-FMLM \cite{chen2019non}, Imputer \citep{chan2020imputer}, and Mask-CTC \citep{higuchi2020mask}. However, during training, one must simulate partial iterations by synthetically masking ground truths or samples from an expert; during inference, one must predetermine criteria to decide which tokens to keep. These lead to train-test mismatch and thus substandard outputs. An alternative from NMT is \textit{iterative refinement} \citep{lee2018deterministic}, which produces full outputs at each iteration and trains on its own iterates. This reduces mismatch and enables retroactive edits, but still lacks flexibility: every iteration is constrained to an initial length $L$ predicted by the model. Viewing ``length-$L$ output sequences'' as a latent space makes it difficult to correct insertions or deletions, which led them to collapse repeating tokens during inference as a post-processing step.

Hence, we propose \textit{iterative realignment}, where \textit{latent alignments} are refined by the decoder. This avoids length prediction and enables more powerful edits, while preserving the flexibility and reduced train-test mismatch of refinement. Our model, Align-Refine, demonstrates this method in an ASR context. 
We use a Transformer-based encoder, and a non-causal Transformer decoder as the refiner. 
The refiner takes the initial alignment and subsequent refinements as inputs. Both the encoder and refiner are supervised with CTC \citep{graves2006connectionist}, a loss defined between sequences and their latent monotonic alignments. Unlike past methods, Align-Refine requires no token masking or expert policies. Repetitions are still collapsed, but this is now incorporated during training by CTC loss.

We validate our approach on two English ASR benchmarks, improving on state-of-the-art infilling methods in inference time (as measured by real-time factor, or RTF) and/or word error rate (WER). On WSJ we close the gap with an autoregressive baseline at 1/14th the RTF, outperforming Mask-CTC after a single iteration. On LibriSpeech we improve on published (LM-free) non-autoregressive results by 2.1\% WER absolute on test-other at $<$1/4th the effective layers (and thus estimated RTF) of Imputer. Our work shows that iterative realignment is a promising direction for other sequence-to-sequence tasks, such as NMT.

\section{Background}
\subsection{Connectionist Temporal Classification}

CTC \citep{graves2006connectionist} is an approach to learning latent monotonic \textit{alignments} from an input sequence $\bx$ to a shorter output sequence $\by$.
Assume our output tokens $y_i$ are in some alphabet $\mathbb{L}$;
then CTC defines an intermediate alphabet $\mathbb{L}' = \mathbb{L} \cup \{\texttt{\_}\}$, where \texttt{\_} is known as a \textit{blank}.
A \textit{CTC alignment} is reduced to an output sequence by collapsing repeated labels, then removing blanks, e.g., \texttt{AB\_\_BB\_A}\,$\mapsto$\,\texttt{ABBA}.
Then, $p(\by | \bx)$ is calculated by marginalizing over all alignments $\psi(\by)$ mapping to an output $\by$. By assuming alignment labels are conditionally independent, we get
\[
    p(\by | \bx) = \sum_{a \in \psi(\by)} p_\theta(a | \bx) = \sum_{a \in \psi(\by)} \prod_{t=1}^{|\bx|} p_\theta(a_t | \bx),
\]
making $J_{\text{CTC}} = - \log p(\by | \bx)$ differentiable and efficiently computed with dynamic programming.

\subsection{Infilling Approaches}

Infilling methods such as Mask-Predict \cite{ghazvininejad2019mask} only allow the model to condition on a subset of the previous sequence. For each iteration $\by^k$:
\[
p(\mathbf{y}^k | \bx) = \prod_i p(y_i^k | \bx, \by^{k-1} \setminus \by^{k-1}_\text{mask}).
\]
Typically, at each iteration a decreasing proportion of high-confidence tokens are unmasked\footnote{
    In Mask-CTC, this is done by the decoder.
    In Imputer, full encoder passes are required, as there is no decoder.
}, so the full budget of $K$ iterations is always required.
Furthermore, the state-of-the-art methods for English ASR are Mask-CTC \cite{higuchi2020mask} and Imputer \cite{chan2020imputer}, which  enforce the added constraint that no decisions may be reversed at all: $p(y_i^{k} | \bx) = \mathbf{1}[y_i^{k} = y_i^{k-1}] \text{ when } y_i^{k-1} \neq \texttt{<MASK>}$.
We demonstrate Mask-CTC in \Cref{fig:qualitative-wsj}.

\section{Align-Refine}
\begin{figure*}[t!]
    \centering
{\ttfamily\fontsize{8pt}{9pt}\selectfont
\addtolength{\tabcolsep}{-6pt}
\setlength\extrarowheight{2pt}
    \begin{tabu}{r@{\hskip 4pt}@{\hskip 4pt}cccccccccccccccccccccccccccccccccccccccccccccc@{\hskip 4pt}@{\hskip 4pt}ccccccccccccccccccccccccccccccccccccccccccccccccccccccccccccc}
     & \multicolumn{46}{c}{\fontsize{9pt}{9pt}\selectfont\textbf{Mask-CTC}, 5 iterations} & \multicolumn{60}{c}{\fontsize{9pt}{9pt}\selectfont\textbf{Align-Refine (char.)}, up to 5 iterations} \\
    \midrule
        \textbf{Enc}  &•&•&•&F&•&•&•&\kern 0.5em&•&•&S&•&•&•&R&E&•&•&•&L&Y&•&•&E&L&•&•&•&•&\kern 0.5em&T&O&•&•&•&•&L&•&\kern 0.5em&H&•&\kern 0.5em&S&A&I&D     &\_&\_&S&S&A&\_&\_&F&F&F&O&\_&O&D&\kern 0.5em&H&'&S&\kern 0.5em&D&I&R&R&E&C&T&L&L&Y&\kern 0.5em&R&R&E&L&A&T&T&E&D&\kern 0.5em&T&O&O&\kern 0.5em&H&E&A&L&T&\kern 0.5em&H&H&E&\kern 0.5em&S&A&A&I&D&D&\_\\
        \textbf{k=1} &•&•&•&F&•&•&•& &H&•&S&•&D&•&R&E&•&•&H&L&Y&•&•&E&L&•&•&•&D& &T&O&•&H&•&•&L&•& &H&•& &S&A&I&D    &\_&\_&S&S&A&\_&\_& &F&F&O&\_&O&D& &H&\bfr{I}&S& &D&I&R&R&E&C&T&L&L&Y& & &R&E&L&A&T&T&E&D& &T&O& & &H&E&A&L&T& & &H&E& &S&A&A&I&D&\_&\_\\
        \textbf{k=2} &•&•&•&F&•&•&•& &H&A&S&•&D&I&R&E&•&•&H&L&Y&•&R&E&L&A&•&•&D& &T&O&•&H&•&A&L&•& &H&•& &S&A&I&D     &\_&\_&S&S&A&\_&\_& &F&F&O&\_&O&D& &H&I&S& &D&I&R&R&E&C&T&L&L&Y& & &R&E&L&A&T&T&E&D& &T&O& & &H&E&A&L&T&\bfr{H}& &H&E& &S&A&A&I&D&\_&\_\\
        \textbf{k=3} &S&•&•&F&•&•&•& &H&A&S& &D&I&R&E&•&T&H&L&Y&•&R&E&L&A&T&•&D& &T&O&•&H&•&A&L&T& &H&•& &S&A&I&D     &\_&\_&S&S&A&\bfr{I}&\_& &F&F&O&\_&O&D& &H&I&S& &D&I&R&R&E&C&T&L&L&Y& & &R&E&L&A&T&T&E&D& &T&O& & &H&E&A&L&T&H& &H&E& &S&A&A&I&D&\_&\_\\
        \textbf{k=4} &S&•&•&F&O&•&•& &H&A&S& &D&I&R&E&•&T&H&L&Y& &R&E&L&A&T&E&D& &T&O& &H&•&A&L&T& &H&E& &S&A&I&D     & \multicolumn{60}{c}{\textrm{\it (identical, so collapse early)}}\\
        \textbf{k=5} &S&E& &F&O&E&T& &H&A&S& &D&I&R&E&C&T&H&L&Y& &R&E&L&A&T&E&D& &T&O& &H&E&A&L&T& &H&E& &S&A&I&D     & \multicolumn{60}{c}{$\downarrow$} \\
        \textbf{End} &S&E& &F&O&E&T& &H&A&S& &D&I&R&E&C&T&H&L&Y& &R&E&L&A&T&E&D& &T&O& &H&E&A&L&T& &H&E& &S&A&I&D                                                                       & \multicolumn{60}{c}{SAI FOOD HIS DIRECTLY RELATED TO HEALTH HE SAID} \\
        \midrule
        & \multicolumn{106}{c}{\fontsize{9pt}{9pt}\textbf{Reference: } SEAFOOD IS DIRECTLY RELATED TO HEALTH HE SAID}
    \end{tabu}
}\vspace{-5pt}
    \caption{WSJ dev93 utterance as decoded by our models. 
    Mask-CTC's masks are denoted with `•'. Mask-CTC gives \texttt{HEALT} as it has no space for \texttt{H} from the very beginning, and outputs \texttt{DIRECTHLY} as it cannot undo \texttt{HLY}'s emission at iteration $k$ = 1. Meanwhile, Align-Refine makes the mistake \texttt{HEALT} immediately, but corrects it over two steps: at $k=1$ it deletes \texttt{HHE}$\mapsto$\texttt{HE}, and at $k=2$ it sees the new space and inserts \texttt{HEALT}$\mapsto$\texttt{HEALTH}.}
    \label{fig:qualitative-wsj}
    \vspace{-10pt}
\end{figure*}

In contrast to these methods, Align-Refine's \textit{encoder} generates an initial proposal, which is then edited iteratively by a non-causal decoder, the \textit{refiner}. Unlike infilling, however, iterative refinement keeps the proposal fully formed (instead of, e.g., masking all but the most confident tokens). This gives decoding flexibility (as iteration can be stopped at any time) and potential speedups (errors are seen and fixed in parallel; easy utterances are refined in fewer steps).

However, unlike \citet{lee2018deterministic}, our proposals are latent CTC alignments $\ba^{k}_{1:|\bx|}$ rather than outputs $\by^{k}_{1:L}$. In previous work, working at the output sequence
level is another source of irreversibility; one needs $p(L | \bx)$ even before the initial proposal $p_{\text{enc}}(\by^0 | L, \bx)$, either explicitly by sampling length predictions \cite{lee2018deterministic} or implicitly by collapsing CTC outputs before infilling \cite{higuchi2020mask}.
Either way, the decoder cannot fix insertions/deletions. Meanwhile, CTC alignments have fixed length and need only be collapsed at inference; insertions/deletions are handled by placing/replacing blanks and spaces (\Cref{fig:qualitative-wsj}). At a given iteration $k$, we have:
\[
p(\by | \bx) = \Ex_{\ba^{k-1}}\!\left[\!\sum_{\ba^k \in \psi(\by)}\!\!p_{\text{ref}}(\ba^k | \ba^{k-1}, \bx)\right].
\]
This is intractable, so like \citet{lee2018deterministic} we take a deterministic lower bound (the mode $\hat{\ba}^{k-1}$). In our case, the loss is now just $J_{\text{CTC}}(\hat{\ba}^{k-1}, \by; \theta_{\text{ref}},  \theta_{\text{enc}})$. Since we don't know \textit{a priori} which $k$ is final, we apply it to $k=1,\dotsc,K$ (for some hyperparameter $K$) with weights $w_1, \dotsc, w_K$. For $k=0$ we get $J_{\text{CTC}}(\bx, \by; \theta_{\text{enc}})$ with weight $\lambda$. 
In summary, we take the greedy alignment at each iteration and apply the CTC loss, as shown in \Cref{fig:align-refine} for $K=2$.
In practice, we upweight the encoder and first iteration terms, then sum to give the total loss.

\section{Experiments}
\begin{figure*}[t]
    \centering
    {\ttfamily\fontsize{8pt}{9pt}\selectfont
\addtolength{\tabcolsep}{-5.6pt}
\setlength\extrarowheight{2pt}
    \begin{tabu}{r@{\hskip 2pt}@{\hskip 2pt}ccccccccccccccccccccccccccccccccccccccccccccccccccccccccccc}
         & \multicolumn{59}{c}{\fontsize{9pt}{9pt}\selectfont\textbf{Align-Refine (subword)}, up to 5 iterations} \\
    \midrule
        \textbf{Enc} & {} [WHEN & \_ & \_ & \_ & [DI & [DI & \_ & CK & I & \_ & \_ & [CAME & \_ & \_ & \_ & \_ & \_ & [DOWN & \_ & \_ & \_ & \_ & \_ & \_ & \_ & \_ & [HIS & \_ & \_ & \_ & \_ & [A & \_ & \_ & UN & UN & T & T & \_ & \_ & \_ & [S & \_ & \_ & IGHT & \_ & LY & \_ & \_ & [S & [S & LA & \_ & \_ & PP & \_ & T & \_ & [HIM \\
        \textbf{k=1} & {} [WHEN & \_ & \_ & \_ & [DI & [DI & \_ & CK & I & \bfb{E} & \bfb{E} & [CAME & \_ & \_ & \_ & \_ & \_ & [DOWN & \_ & \_ & \_ & \_ & \_ & \_ & \_ & \_ & [HIS & \_ & \_ & \_ & \_ & [A & \_ & \_ & \textbf{\_} & UN & T & T & \_ & \_ & \_ & [S & \bfb{L} & \_ & IGHT & \_ & LY & \_ & \_ & \textbf{\_} & [S & LA & \_ & \_ & PP & \_ & \bfb{ED} & \_ & [HIM \\
        \textbf{k=2} & {} [WHEN & \_ & \_ & \_ & [DI & [DI & \_ & CK & I & \textbf{\_} & E & [CAME & \_ & \_ & \_ & \_ & \_ & [DOWN & \_ & \_ & \_ & \_ & \_ & \_ & \_ & \_ & [HIS & \_ & \_ & \_ & \_ & [A & \_ & \_ & \_ & UN & \bfr{\_} & T & \_ & \_ & \_ & [S & L & \_ & IGHT & \_ & LY & \_ & \_ & \_ & [S & LA & \_ & \_ & PP & \_ & ED & \_ & [HIM \\
        \textbf{k=3} & \multicolumn{59}{c}{\textrm{\it (identical, so collapse early)}} \\
        \textbf{...} & \multicolumn{59}{c}{$\downarrow$} \\
        \textbf{End} & \multicolumn{59}{c}{[WHEN\;\;[DICKIE\;\;[CAME\;\;[DOWN\;\;[HIS\;\;[AUNT\;\;[SLIGHTLY\;\;[SLAPPED\;\;[HIM}
    \end{tabu}
}    
    \caption{LibriSpeech test-other utterance; reference matches the prediction. At $k$ = 1 three separate corrections are made, two of which (\texttt{DICKI}$\mapsto$\texttt{DICKIE}; \texttt{SLAPPT}$\mapsto$\texttt{SLAPPED}) cannot be done from the audio. In $k$ = 2 the multimodal prediction \texttt{E\,E} is resolved into \texttt{\_\,E}, though the repetition-collapsed transcript would be correct regardless. }
    \label{fig:qualitative-librispeech}
    \vspace{-7.5pt}
\end{figure*}

(See \Cref{sec:training} for additional details.)

\paragraph{Data.} 
We evaluate on two English ASR benchmarks: WSJ (81 hours; \citealp{paul1992wsj}) and LibriSpeech (960 hours; \citealp{panayotov2015librispeech}).
For WSJ we run at the character level to match Mask-CTC; for LibriSpeech, we build a 400-token BPE vocabulary to match Imputer. We use 80-dim.\ filter banks and SpecAugment \citep{park2019specaugment}. 

\paragraph{Model.}
We use a 12-layer encoder and 6-layer decoder, similar to Mask-CTC \citep{higuchi2020mask}; each layer has 4 heads of 256 units. For LibriSpeech we use 8 heads of 512 units.
We take $K$ = 4 unless stated otherwise.

\paragraph{Decoding.}
We evaluate with decoding iterations $k$ from \{0, 1, 3, 5, 10\}. In each case we exit early if consecutive iterations are identical. The final CTC alignment is collapsed to give the result. To match previous non-autoregressive ASR work, we do not use a language model (LM).

\paragraph{WSJ results (\Cref{tab:wsj}).} Mask-CTC and Align-Refine have the same architecture; the difference is in training and evaluation. Joint training with the refiner improves the encoder as a standalone CTC model ($k = 0$) to a similar degree. However, Align-Refine's benefit is clear by $k = 1$, where it reduces the initial alignment's WER by 1.9\% absolute and outperforms Mask-CTC at any \# of iterations:

\begin{table}[!ht]
    \centering
    \small
    \setlength{\tabcolsep}{3pt}
    \vspace{-5pt}
   \begin{tabular}{@{}lccccl@{}}
        \toprule
                                                & \multicolumn{2}{c}{\textbf{\# passes}} & \multicolumn{2}{c}{\textbf{WER}} \\
         \textbf{Model}                                            & \textbf{Enc} & \textbf{Dec} & \textbf{dev93} & \textbf{eval92} & \textbf{RTF} \\
         \midrule
         \multicolumn{5}{@{}l}{\textit{Autoregressive baseline} \cite{higuchi2020mask}}                            &  \\
         \;\;\;\textsc{CTC+Attention}              & 1 & $L$ & 14.4 & 11.3  & 0.97* \\
         \;\;\;\;\;\; + beam search                         & 1 & $>$\,$L$ & 13.5 & 10.9  & 4.62* \\
         \midrule               
         \multicolumn{6}{@{}l}{\textit{Previous work} \cite{higuchi2020mask, chan2020imputer}}                       \\
         \;\;\;\textsc{CTC}         & 1 & -- & 22.2 & 17.9 & 0.03* \\
         \;\;\;\textsc{Mask-CTC}    & 1 & 0 & 16.3 & 12.9 & 0.03* \\
             & 1 & 1 & 15.7 & 12.5 & 0.04* \\
             & 1 & 5 & 15.5 & 12.2 & 0.05* \\
             & 1 & \#mask & 15.4 & 12.1 & 0.13* \\
         \;\;\;\textsc{Imputer (8-lyr)} \textdagger    & 8 & -- & -- & 12.7 & -- \\
         \midrule
         \textit{Our work}                                &  & &      & \\
         \;\;\;\textsc{CTC}                        & 1 & -- & 18.6 & 15.0                  & 0.036 \\
         \;\;\;\textsc{Mask-CTC}         & 1 & 5 & 15.3 & 12.8                  & 0.072 \\
         \;\;\;\textsc{Align-Refine}               & 1 & 0 & 16.2 & 13.5 & 0.037 \\
                        & 1 & 1 & 14.1 & 11.6 & 0.048 \\
                        & 1 & 3 & 13.9 & 11.5 & 0.066 \\
                        & \textbf{1} & \textbf{5} & \textbf{13.7} & \textbf{11.4} & \textbf{0.068} \\
                        & 1 & 10 & 13.7 & 11.4 & 0.068 \\
         \bottomrule
    \end{tabular}
    \caption{
        Non-autoregressive ASR on WSJ. No LMs are used. *: RTFs from \citeauthor{higuchi2020mask}, which are lower than ours for corresponding models. \textdagger: No SpecAugment.
    }
    \vspace{-5pt}
    \label{tab:wsj}
\end{table}

\noindent By $k=5$, Align-Refine closes the performance gap with a comparable autoregressive model at 1/14th the RTF. As for the 8-layer Imputer, it seems unlikely that re-introducing SpecAugment would outperform this augmented 12+6-layer autoregressive baseline; even if performances did match, RTFs would be higher than Align-Refine's (\Cref{sec:discussion}).

\paragraph{LibriSpeech results (\Cref{tab:librispeech}).} Align-Refine gives 9.0\% WER on test-other with no LM, outperforming published non-autoregressive models by 2.1\% WER absolute. This is 5.1 points absolute over training the encoder with CTC only, even after SpecAugment is used (compare with 1.9 points from 16-layer CTC to Imputer). In \Cref{fig:qualitative-librispeech} we see the refiner make output-conditional edits that would be difficult for greedy CTC inference; we attribute our outsized gain on test-other to this LM-like behavior. Future work could start from stronger CTC encoders like QuartzNet \cite{kriman2020quartz} to achieve even better results.

\begin{table}[!ht]
    \centering
    \small
    \setlength{\tabcolsep}{3pt}
    \begin{tabular}{@{}lcccc@{}}
        \toprule
          & \multicolumn{2}{c}{\textbf{\# passes}}  & \multicolumn{2}{c}{\textbf{WER (test)}} \\
         \textbf{Model}                                            & \textbf{Enc} & \textbf{Dec} & \textbf{clean} & \textbf{other} \\
         \midrule
\multicolumn{5}{@{}l}{\textit{Autoregressive models} \citep{han2020context}} \\
         \;\;\;\textsc{LAS (360M)}            & 1 & $L$ & 2.6 & 6.0  \\
         \;\;\;\textsc{RNN-T (ContextNet-L)}            & 1 & $L$  & 2.1 & 4.6  \\
         \midrule
\multicolumn{5}{@{}l}{\textit{Previous work} \citep{chan2020imputer, kriman2020quartz}} \\
         \;\;\;\textsc{CTC (16-lyr)} \textdagger       & 1 & -- & 4.6 & 13.0 \\
         \;\;\;\textsc{Imputer (16-lyr)} \textdagger     & 8 & -- & 4.0 & 11.1 \\
         \;\;\;\textsc{CTC (Jasper DR 10x5)}    & 1 & -- & 4.3 & 11.8 \\
         \;\;\;\textsc{CTC (QuartzNet 15x5)}    & 1 & -- & 3.9 & 11.2 \\
         \midrule
         \textit{Our work}                                         &   &      & \\
         \;\;\;\textsc{CTC}          & 1 & -- & 5.1 & 14.1                  \\
         \;\;\;\textsc{Align-Refine}     & 1 & 0 & 4.6 & 11.5 \\
              & 1 & 1 & 3.8 & 9.5 \\
              & \textbf{1} & \textbf{3} & \textbf{3.6} & \textbf{9.0} \\
              & 1 & 5 & 3.6 & 9.0 \\
         \bottomrule
    \end{tabular}
    \caption{
        Non-autoregressive ASR on LibriSpeech. No LMs are used.  \textdagger: No SpecAugment.
            }
    \vspace{-15pt}
    \label{tab:librispeech}
\end{table}

\section{Discussion}
\label{sec:discussion}

In all, we have seen that Align-Refine improves performance over infilling methods like Imputer and Mask-CTC across tokenizations and dataset sizes (Tables \ref{tab:wsj} and \ref{tab:librispeech}). We saw improvements qualitatively as parallel and multi-stage insertion/deletion edits (Figures \ref{fig:qualitative-wsj} and \ref{fig:qualitative-librispeech}). Some final considerations:

\paragraph{Speed.} One advantage of Align-Refine over the encoder-only Imputer is that it factors out refinement from feature processing, allowing for adaptive decoding times. \citet{kasai2020deep} critiqued this separation in non-autoregressive models by showing autoregressive models could shift layers from decoder to encoder to reduce the speed gap;
however, we show that Align-Refine benefits from a similar reallocation (\Cref{tab:depth-ablation}):
\begin{table}[h]
    \centering
    \small
   \begin{tabular}{@{}XlcccccX@{}}
        \toprule
         & \multicolumn{5}{c}{\textbf{WER (test-other) for each $k$}} \\
         \textbf{Model-(\# enc)-(\# dec)} & \textbf{0}        & \textbf{1}     & \textbf{2}     & \textbf{3}     & \textbf{5}    \\
         \midrule
         \textsc{Align-Refine-12-6}                   & 12.5     & 10.0  & 9.4  & 9.3 & 9.3      \\
         \textsc{Align-Refine-15-3}                     & 10.5     & 9.5   & 9.4   & 9.3   & 9.3   \\
         \textsc{Align-Refine-17-1}                     & 10.4        & 10.0     & 10.0     & 10.0     & 10.0     \\
         \bottomrule
    \end{tabular}
    \caption{
        Results with various encoder-decoder splits on LibriSpeech. We take $K=2$ to speed up training.
    }
    \vspace{-5pt}
    \label{tab:depth-ablation}
\end{table}

\noindent The refiner should have some depth: 17-1 underperforms on test-other (though within 0.1 points on test-clean).
However, 15-3 performs as well as 12-6 at $k=1$ onwards, despite passing through half the number of decoder layers. At $k=3$, our LibriSpeech model's RTF was 0.171, while 15-3's RTF is 0.136.
In this configuration, we pass through 15 encoder and at most 3 $\times$ 3 = 9 decoder layers. To put this in context, each Imputer inference passes through 16 $\times$ 8 = 128 encoder layers with the same alignment-length inputs and 8 head, 512 unit size.

\paragraph{Limitations.} The refines sometimes makes edits which do not affect post-collapse outputs (\Cref{fig:qualitative-librispeech}, $k=2$); variants that use repetition tokens (ASG; \citealp{collobert2016wav}) or prohibit repetition collapse \citep{chan2020imputer} may mitigate this behavior.
The initial downsampling also restricts what edits can be done in one step (Figure \ref{fig:qualitative-wsj}, $k=1,2$).
Finally, word-level errors like \texttt{SAI}\;\texttt{FOOD}\;\texttt{HIS} may require new mechanisms to fix;
future work could integrate non-autoregressive LMs like BERT \citep{devlin2019bert} via alignment synthesis, fusion, or otherwise (cf.\ \citealp{salazar2020mlm}).

\section*{Acknowledgements}
We are grateful to the AWS Speech Science team and the Stanford NLP Group for their advice and feedback, and to Yosuke Higuchi and William Chan for their helpful correspondence.

\bibliography{acl2020}
\bibliographystyle{acl_natbib}

\appendix

\section{Training Details}
\label{sec:training}
Our setup extends the Mask-CTC recipe (\url{https://github.com/espnet/espnet/pull/2223/}) in ESPnet \citep{watanabe2018espnet}. We will release our Align-Refine recipe and ESPnet code changes by publication time at \url{https://github.com/amazon-research/}.

\paragraph{Architecture.} Following Mask-CTC, the 12-layer encoder has a convolutional frontend that downsamples input lengths and features by 4x, using two layers of 2D convolutions of filter size 3x3 and stride 2 \cite{dong2018speechtrans}.
The 6-layer decoder has no causal attention masks. Pre-norm residual paths are used \citep{nguyen2019transformers}.

\paragraph{Training.} We apply dropout with $p=0.1$ on WSJ and $p=0.2$ on LibriSpeech. The initial CTC is weighted with $\lambda = 0.3$; remaining weights are spread over the $K$ iterations, with $w_1$ three times larger than the rest.
For WSJ, we used a batch size of 32 sequences.
For LibriSpeech, we used a batch size of 25000 tokens.
Gradients are accumulated in both cases. LibriSpeech models are trained over four V100 GPUs (p3.8xlarge instances on AWS). We trained to convergence on WSJ, and for 125K steps on LibriSpeech.
Following Mask-CTC, we use label smoothing of 0.1 and the standard Transformer inverse square-root learning rate schedule with a linear warmup of 25000 steps. We double their \texttt{transformer-lr} factor to 10.0.

\paragraph{Decoding} At decode time, we average model weights over 30 training checkpoints.
All RTFs are measured on a single CPU thread (\texttt{-{}-nj 1, ngpu=0}). We take real-time duration (\texttt{time}) and divide by the total utterance duration in ESPnet's \texttt{utt2dur} file on \texttt{eval92} for WSJ and \texttt{test-other} for LibriSpeech.

\end{document}